%% file: adblock.tex
\documentclass[sigconf]{acmart}

\usepackage{helvet}
\usepackage{courier}
\usepackage{amsmath}
\usepackage{graphicx}
\usepackage{url}
\usepackage[font={small}]{subfig}
\usepackage[show]{chato-notes}

\newcommand{\spara}[1]{\smallskip\noindent{\bf #1}}

\newcommand{\homepages}{{\tt Top150}}
\newcommand{\gdeltdata}{{\tt GDELT}}

\begin{document}
%

\copyrightyear{2017} 
\acmYear{2017} 
\setcopyright{acmlicensed}
\acmConference{WebSci '17}{June 25-28, 2017}{Troy, NY, USA}\acmPrice{15.00}\acmDOI{http://dx.doi.org/10.1145/3091478.3091514}
\acmISBN{978-1-4503-4896-6/17/06}

\title{Ad-blocking: A Study on Performance, Privacy and Counter-measures}

\author{Kiran Garimella}
\affiliation{%
  \institution{Aalto University}
  \city{Helsinki} 
  \country{Finland} 
}
\email{kiran.garimella@aalto.fi}

\author{Orestis Kostakis}
\affiliation{%
  \institution{Aalto University}
  \city{Helsinki} 
  \country{Finland} 
}
\email{orestis.kostakis@aalto.fi}

\author{Michael Mathioudakis}
\affiliation{%
  \institution{Aalto University}
  \city{Helsinki} 
  \country{Finland} 
}
\email{michael.mathioudakis@aalto.fi}

\renewcommand{\shortauthors}{Garimella, Kostakis, Mathioudakis}

\begin{abstract}
Many internet ventures rely on advertising for their revenue. However, users feel discontent by the presence of ads on the websites they visit, as the data-size of ads is often comparable to that of the actual content. This has an impact not only on the loading time of webpages, but also on the internet bill of the user in some cases. In absence of a mutually-agreed procedure for opting out of advertisements, many users resort to ad-blocking browser-extensions. 

In this work, we study the performance of popular ad-blockers on a large set of news websites. 
Moreover, we investigate the benefits of ad-blockers on user privacy as well as the mechanisms used by websites to counter them. Finally, we explore the traffic overhead due to the ad-blockers themselves.
\end{abstract}

\maketitle

\input{introduction.tex}
\input{related.tex}

\input{dataset.tex}
\input{findings.tex}

\bibliographystyle{ACM-Reference-Format}
\bibliography{related}
\end{document}

%% file: introduction.tex
\section{Introduction}

Ad-blocking is now a widespread practice among web users.
According to current estimates, it is employed by around 200 million desktop and 300 million mobile users, a user base that still grows at 40\% annually~\cite{pagefair2016report}.
The same estimates indicate that ad-blocking translates into significant losses for publishers, reaching 41 billion US dollars for 2016.
It is therefore not surprising that there is an arms race between publishers and advertisers, on one hand, and ad-blocking tools (or `ad-blockers'), on the other\footnote{{https://techcrunch.com/2016/08/11/facebooks-blocker-blocking-ads-blocked-by-blockers/}}.
In this context, it is clear that understanding the mechanisms implemented by ad-blockers and their effects is of wide and continuous interest.

In this paper, we study a number of popular ad-blockers and analyze their performance in desktop and mobile settings for a large number of webpages. 
Specifically, we employ the ad-blockers on two sets of webpages: one consisting of the front pages of popular news websites; and another consisting of around 30,000 web articles. 

For each webpage, we collect a number of measures that describe the browser workload with and without the use of an ad-blocker -- including, for instance, the amount of data loaded, the load-time of webpages, or the number of data requests.
Our results indicate that ad-blockers decrease the data consumption significantly, though the benefits in terms of load times are limited.
Moreover, by taking a deeper look into the set of requests incurred by websites and ad-blockers, we identify cases where the websites attempt to counter the ad-blockers, but also requests that the ad-blockers incur on their own.
Finally, we investigate the extent to which ad-blockers prevent the transfer of user-tracking information, resulting in privacy benefits for users. 
When comparing a range of different ad-blockers, we find that uBlock performs the best, in terms of both data savings and user-tracking.




%% file: related.tex
\section{Related work}

\textbf{Online Advertising}
Advertising has emerged as one of the main sources of revenue on the web \cite{evans2008economics}. 
On one hand, advertisers claim that ads on the web help keep most parts of the web free for consumers \cite{barford2014adscape}.
On the other hand, consumers find advertising annoying and obstructive. In one study, the New York Times analyzed the top 50 news websites\footnote{\url{http://www.nytimes.com/interactive/2015/10/01/business/cost-of-mobile-ads.html}} and found that more than half of all data came from ads.

Moreover, many users perceive web ads as a threat to privacy \cite{falahrastegar2014anatomy} and tracking is cited as one of the most common reasons for users to install ad-blockers~\cite{pagefair2015report}.
This is corroborated by many studies that show large scale tracking behavior on the web \cite{falahrastegar2014anatomy,schelter2016ubiquity}.
For instance, Yu et al. \cite{yu2016tracking} find that 95\% of the pages visited contain 3rd party requests to potential trackers and 78\% attempt to transfer unsafe data.
To counter tracking, ad-blockers also remove tracking buttons (such as Facebook's `like' button) and protect their users from known malware domains.


\textbf{Ad blocking}
In~\cite{malloy2016ad}, the authors propose methods to detect the users who have installed an ad-blocker and characterize ad-block usage for a large set (2 millions) of users. In addition, they give details on demographics and geographic penetration of ad-blockers on the web.
In \cite{pujol2015annoyed}, the authors analyze the use of two ad-blockers, AdblockPlus and Ghostery, on web traffic data from a European ISP. 
They show that there is a drop in requests to third party services when using ad-block and estimate that around 18\% of the traffic is due to ads.
There are a few differences between our study and theirs: (i) their study handles only two ad-blockers, (ii) with a constant battle between advertisers and ad-blockers, the ad-blocking and counter ad-blocking scene has been changing over the last few years, and so have some of the related performance measurements. (iii) our study handles in-html ads, which are used by advertisers nowadays to bypass ad-blocking.

In \cite{wills2016ad}, the authors classify third party trackers into various categories including ad trackers, analytics, social, etc and compare the performance of various ad-blocking tools with respect to blocking these third party requests. They find that there is a lot of variance in the type of requests that are blocked by each ad-blocker. They then look deeper into the individual default and possible configurations of these ad-blockers and study the changes in blocking capabilities with the different settings. 

There have been privacy concerns with ad-blockers too. Many popular ad-blockers (including AdBlock and AdblockPlus) participate in the {\it Acceptable Ads} program\footnote{\url{https://acceptableads.com/}}, allowing `non-intrusive' ads to go through.
For an extensive study on the issue see \cite{walls2015measuring}.

To counter the effects of increasing ad-blocking, advertisers are relying on counter-ad-blocking tools. In~\cite{nithyanand2016ad}, the authors study their use on the most popular five thousand websites and find that at least 6.7\% of these sites use them.


%% file: dataset.tex
\section{Data}
We analyze two datasets, chosen to capture potential differences between content-heavy homepages of large websites and individual webpages.

\spara{\homepages} This is a list of the top 150 news websites, as ranked by Alexa\footnote{\url{http://www.alexa.com/topsites/category/News/Newspapers}. We processed the list manually and removed a few which were not standard news sites (e.g. reddit.com)}.
For each of them, we have a URL that points to the website homepage, for both its  desktop and mobile version (unless the latter does not exist).

\spara{\gdeltdata} A list of 30,000 URLs pointing to different news articles published on a single day  (November 8, 2016). The list was obtained from GDELT\footnote{\url{http://gdeltproject.org/}}, a project that collects news stories from all around the world over the years.

\smallskip

We load the webpage of each URL with a clean instance of Chrome browser (no cached content or extensions), using the Selenium Python library\footnote{\url{http://selenium-python.readthedocs.io/}} on a Macbook pro with 8 cores, with no other major processes running.
On each load, we capture the HAR file for the load.
The HAR (\underline{H}TTP \underline{A}rchive \underline{F}ile) is a JSON-formatted file that captures the interactions between the browser and the website, including network requests, types and size of objects, and load times.

We load the same URL in six browser modes, all simultaneously: a vanilla mode (no ad-blocker), and one mode for each of five ad-blockers.
The ad-blockers are AdBlock\footnote{\small{\url{https://getadblock.com/}}}, AdblockPlus\footnote{\small{\url{https://adblockplus.org}}}, Ghostery\footnote{\small{\url{https://www.ghostery.com/}}}, uBlock\footnote{\small{\url{https://www.ublock.org}}} and Privacy Badger\footnote{\small{\url{https://www.eff.org/privacybadger}}} -- chosen from the most popular ad-blockers on the Chrome Store.
For the \homepages\ dataset, we also load the mobile version of the page (as loaded on Google Nexus 5) using Selenium's mobile emulation tool\footnote{\url{https://sites.google.com/a/chromium.org/chromedriver/mobile-emulation}}. Note that, though we report the data about mobile websites with and without ad-block, the ad-blockers used were for the desktop version. There are no ad-blockers for the mobile version of Chrome and none for any of the Android browsers.
%
Our datasets and code can be accessed here\footnote{\url{https://users.ics.aalto.fi/kiran/adblock/}}.

%% file: findings.tex
\section{Findings}

This section details our findings on the efficiency of ad-blockers, the ad-blocker benefits on user privacy, the counter-measures used by advertisers, and finally the traffic load that ad-blockers incur on their own.

\subsection{Ad-blocker Efficiency}

For each URL and setting (desktop or mobile, with or without ad-blocker), we collect measures that describe the average performance of the browser in loading the webpages of a dataset. The measures include: the number of distinct domains for all HTTP requests performed during loading; the maximum number of threads that run concurrently at any point; the total number of HTTP requests; the total size of downloaded and uploaded data; the cumulative time for loading the webpage, obtained by summing up the durations of all requests; and, finally, the wall-clock time.
We acquire these statistics by parsing the HAR files.

The number of distinct domains indicate the number of different parties that acquire information of the user, such as IP address and the user-agent string (contains device type, name and version of browser and OS, etc).
The number of HTTP requests and maximum number of concurrent threads give an indication of the load on the user's machine. 
The amount of transferred data, and the cumulative and wall-clock time affect the user experience.

\begin{figure}
\centering
\includegraphics[width=0.9\columnwidth]{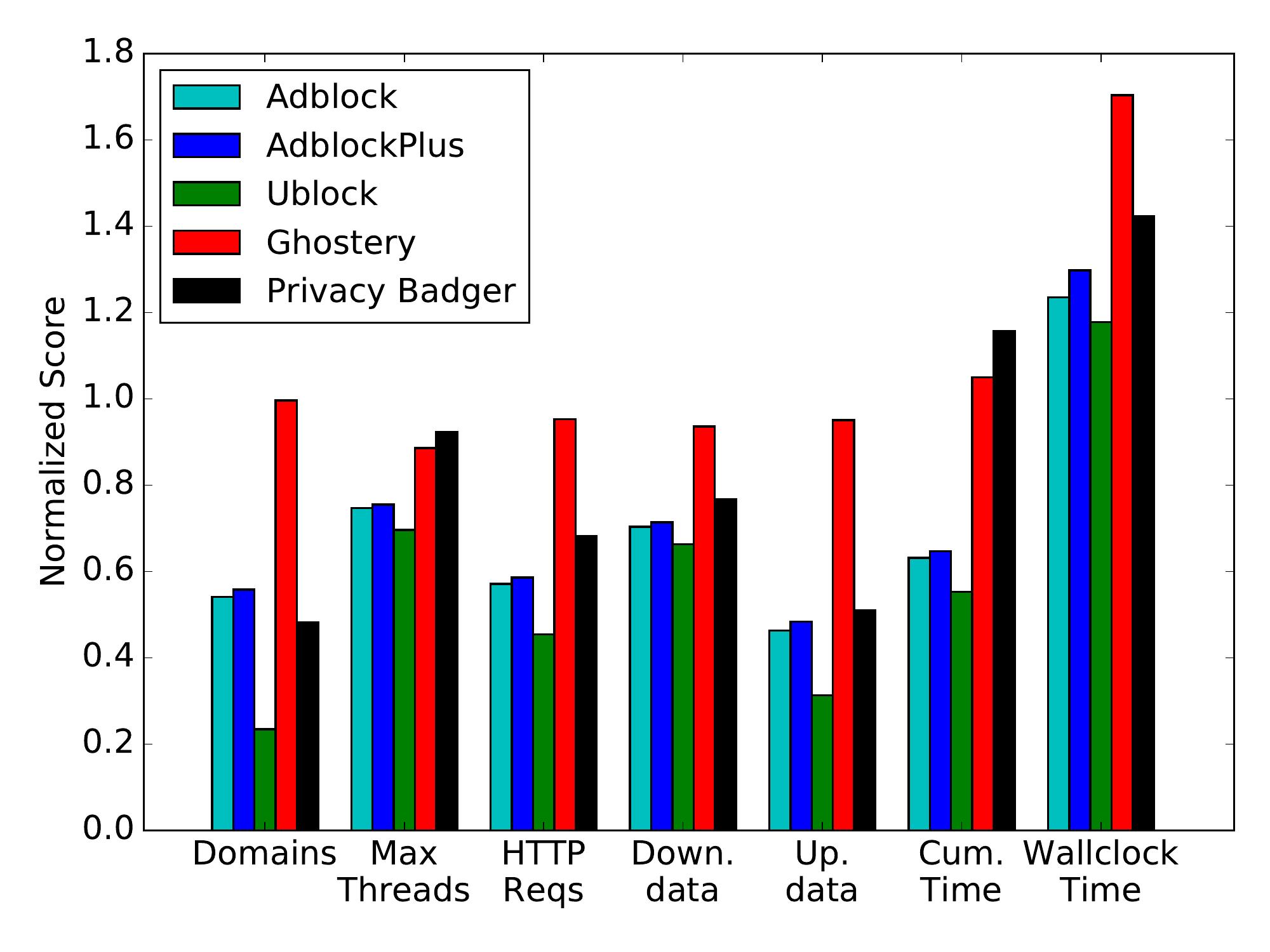}
\caption{Performance benchmark  when using an ad-blocker on the desktop version of \homepages, normalized over the vanilla mode.}%
\label{fig:basicstats150}
\vspace{-\baselineskip}
\end{figure}
\begin{figure}
\centering
\includegraphics[width=0.9\columnwidth]{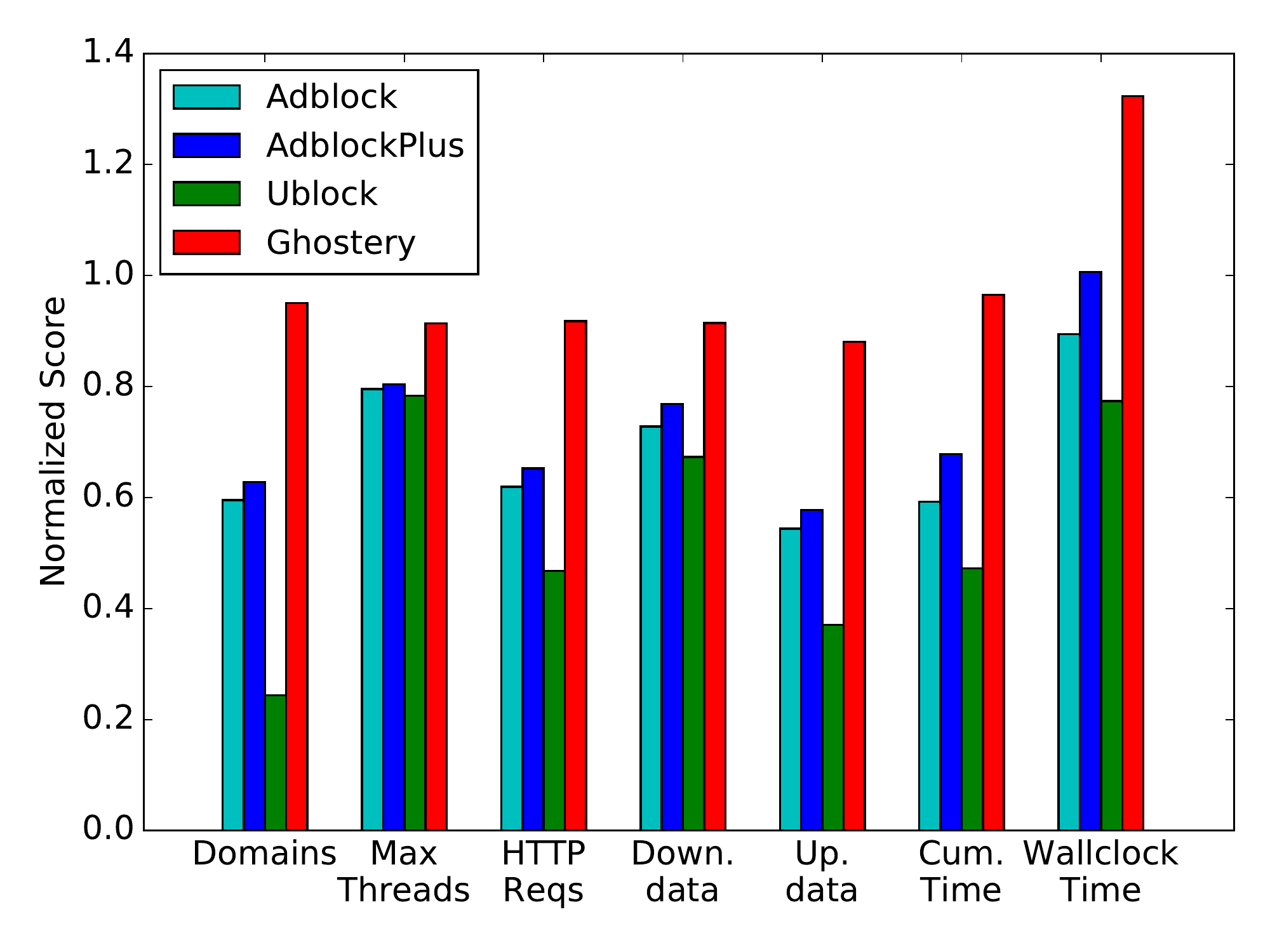}
\caption{Performance benchmark when using an ad-blocker on the mobile version of \homepages, normalized over the vanilla mode. Privacy Badger was discarded because of inconsistencies in the data collection.}%
\label{fig:basicstats150mobile}
\vspace{-\baselineskip}
\end{figure}
\begin{figure}[t]
\centering
\includegraphics[width=0.9\columnwidth]{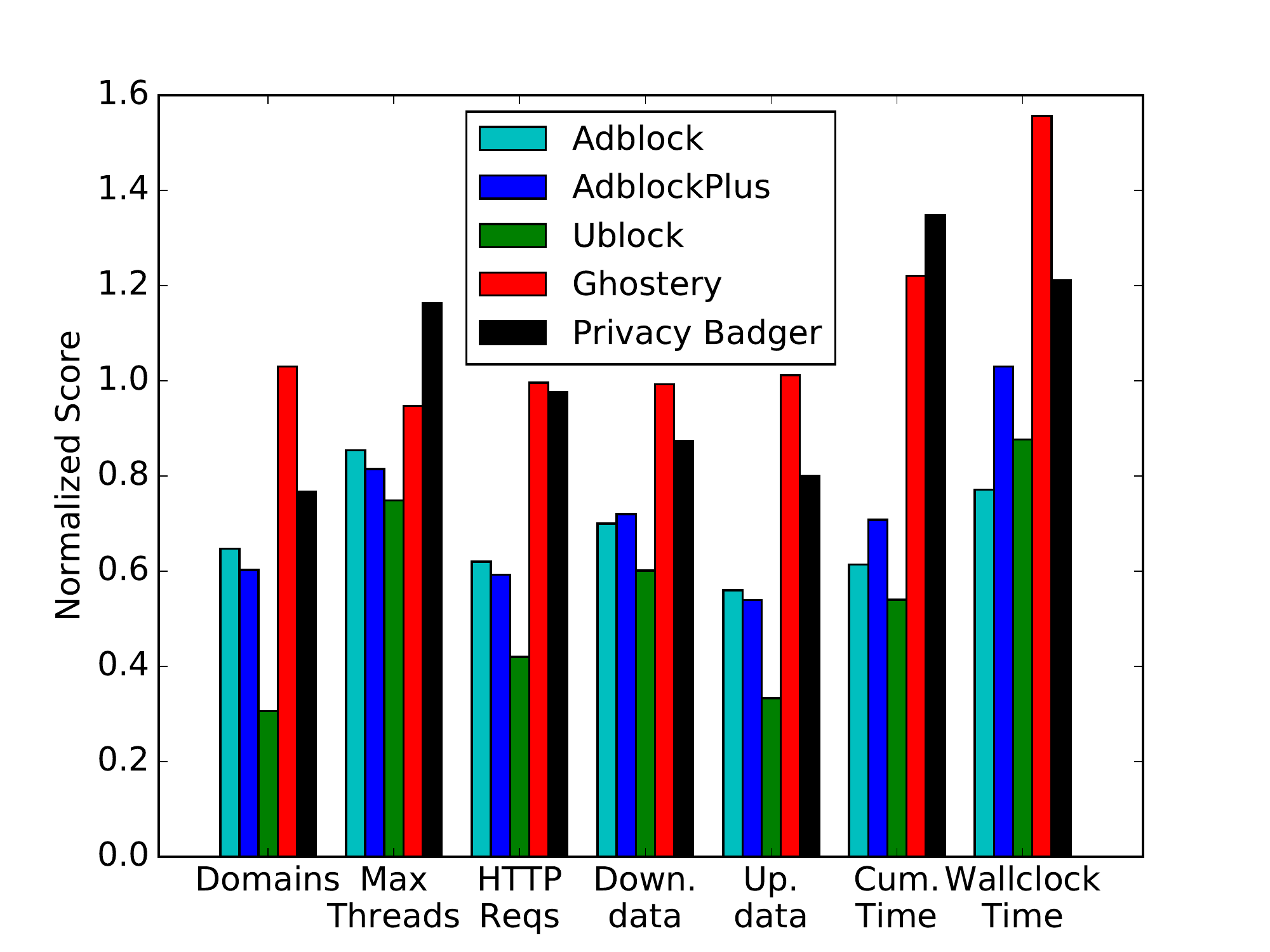}
\caption{Performance measures when using an ad-blocker on the \gdeltdata\ dataset, normalized over the vanilla mode.}%
\label{fig:basicstatslarge}%
\vspace{-\baselineskip}
\end{figure}

Figures \ref{fig:basicstats150} and \ref{fig:basicstats150mobile} show the results for the desktop and mobile versions of the \homepages\ webpages; and Figure~\ref{fig:basicstatslarge} for the \gdeltdata\ webpages.
Numbers represent the ratio of the measures over the vanilla mode -- for example, a value of 0.6 for the number of domains means that using an ad-blocker loads 60\% the number of domains compared to using no ad-blocker.

We find that: (i) All ad-blockers except Ghostery give around 25-34\% savings in the amount of data transferred (on average).
This is a bit higher than the 18\% saving reported by \cite{pujol2015annoyed} and 13-34\% reported by \cite{wills2016ad}.
Ghostery is the exception because it is not an ad-blocking tool per se, but gives a choice for a user to track who is tracking them and block those of their choice. So by default Ghostery does not block any content. We still include it in all our measurements, as it still makes sense to study it in other analysis cases that follow in this section.
(ii) On average, there is an \emph{increase} in the wall clock time when using an ad-blocker, even as the cumulative time decreases for some of the ad-blockers. This means that ad-blockers incur overhead, but block ad-related threads that were meant to run in parallel while loading content; this overhead is not necessarily experienced by the user, since various scripts can often run in the background, after the important content has loaded.
(iii) uBlock gives the best performance, both in terms of data and time saved. 

\subsection{Benefits for User Privacy}
By examining the HTTP requests issued by the browser when loading a webpage, we notice that a significant amount of these are intended for tracking the users' behaviour.
We attempt to quantify the potential privacy hazard for the user when accessing a webpage by analyzing the parameters passed through these requests.
Specifically, to measure the impact of potentially privacy invading requests when using no ad-blocker, we manually examined the tracking parameters from the request headers and found nine frequently used tracking parameters that accompanied a large set of requests.
The parameters were: `pixel', 'track', `google\_gid', `user\_id', `partnerid', `partnerUID', `partner\_device\_id', `uid', `user\_cookie'.
The presence of these parameters offers a high-precision method to identify user-tracking, although we currently do not quantify its recall.

\begin{figure}
\centering
\includegraphics[width=0.9\columnwidth]{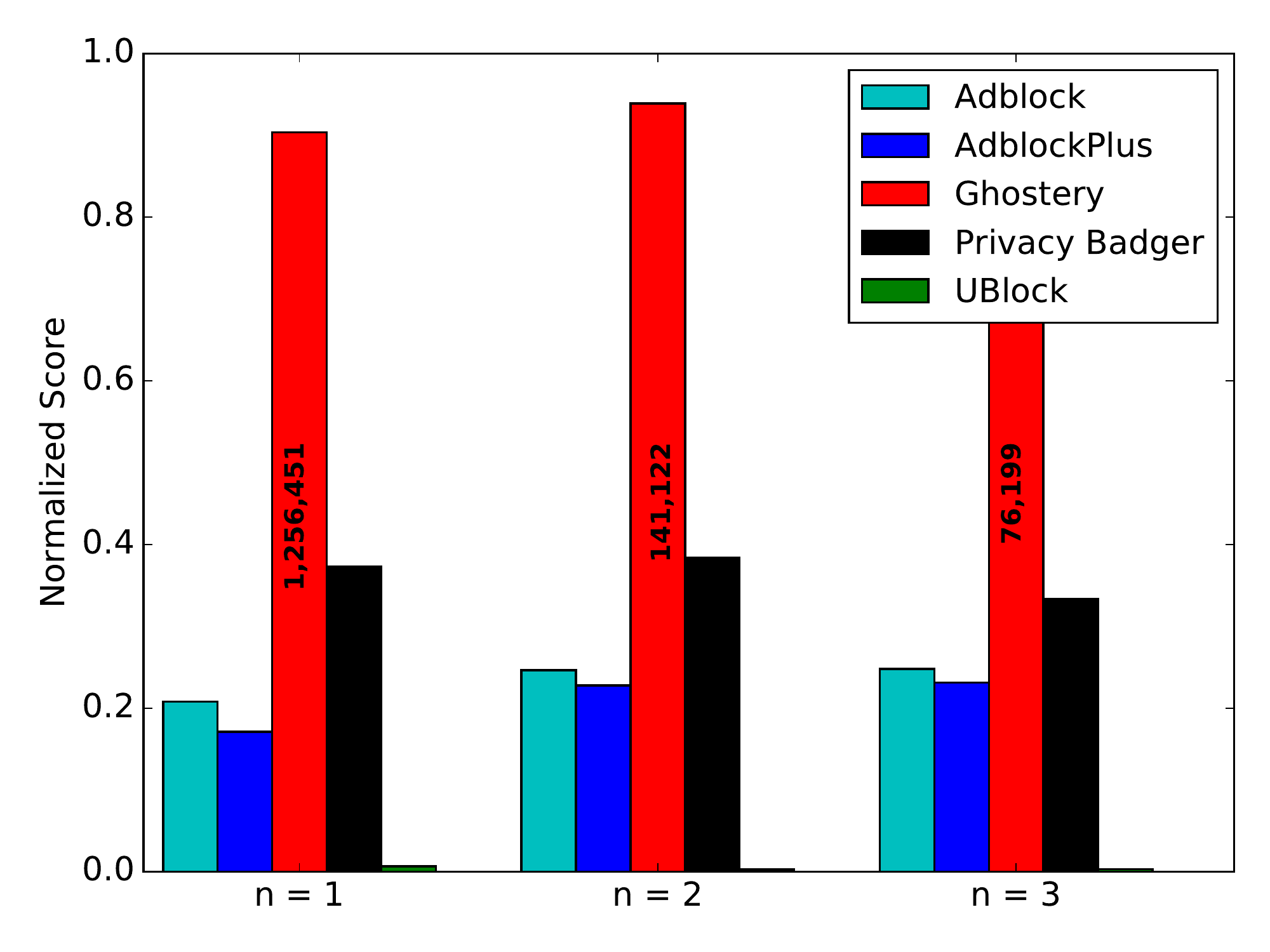}
\caption{Fraction of distinct requests containing tracking keywords for one, two and three tracking parameters. The black text in the red bar shows the total number of requests containing these tracking parameters in the vanilla mode.}
\label{fig:privacystats}
\vspace{-\baselineskip}
\end{figure}

Figure \ref{fig:privacystats} shows the fraction of requests containing one, two and three tracking parameters in a request when using an ad-blocker, again normalized over the vanilla mode.
We can see from the figure that, again, uBlock achieves near-perfect performance, whereas other ad-blockers (except Ghostery) block about 60-80\% of requests with such parameters. The performance remains the same, when considering requests with different numbers of tracking parameters, though the total number of requests goes down exponentially.



\subsection{Blocking the Ad-blockers}

Online ventures relying on advertising try to counter to the ad-blocking tools \cite{nithyanand2016ad}.
In our analysis, we witness several techniques for countering ad-blockers, with different goals.
In certain cases, the goal is simply to detect whether the browser has an active ad-blocking extension.
Upon detection, a message is rendered on the webpage and notifies the user about the tool.
This message may simply notify the user that the tool interferes with the content and the user-experience.
Other times, the message blocks the user from accessing the actual content, until they have turned off the tool.
In more extreme cases, the goal is to circumvent the ad-blocking tool altogether. 

We examine again the HTTP requests performed when loading a webpage and  look for specific Javascript modules designed to counter ad-blocking, such as 
\texttt{blockadblock.js}, \texttt{inn-anti- adblock}, and \texttt{adblock html cache buster}.
When webpages from the \gdeltdata\ dataset are loaded in the vanilla mode (no ad-blockers), 
we find more than 3,260 webpages ($10.8\%$) with explicit attempts
 to determine if any ad-blocking tool is used.
When \texttt{AdBlock} or \texttt{AdblockPlus} is activated, the number of webpages rises to 10,097 and 10,226 $(33\%)$, respectively, indicating that a large portion of webpages actively track the usage of these ad-blockers.
When the other tools are used, the number of webpages with such requests is not as high; $15,9,8\%$ of the webpages for \texttt{Ghostery}, \texttt{Privacy Badger}, and \texttt{uBlock} respectively.
The variation in numbers between AdBlock and AdblockPlus, on one hand, and the rest of the ad-blockers, on the other, can be explained by the popularity of the former -- but also the effectiveness of the ad-blockers in blocking such requests. 

\subsection{Requests due to Ad-blockers}

In this section, we try to dissect what \emph{additional} requests are made when using an ad-blocker.
Consider the set of HTTP requests required to render a webpage in the vanilla mode, denoted by $\mathcal{V}$.
Let $\mathcal{B}$ be the set of requests when using an ad-blocking tool,
then we define 
$\mathcal{A}=\mathcal{V}\setminus\mathcal{B}$ as the identified advertisements,
and  $\mathcal{C}=\mathcal{V}\cap\mathcal{B}$ as the ``true'' content of the page.
Finally,  $\mathcal{E}=\mathcal{B}\setminus\mathcal{V}$ is ``extra'' content loaded due to or by the ad-blocking tool.
Unsurprisingly, the most popular domains found in $\mathcal{A}$ are
\texttt{googlesyndication.com} and \texttt{doubleclick.net}, in accordance to previous studies, and blocked (characterized as ads) by all tools.

A more interesting question is to identify the ``extra'' $\mathcal{E}$ requests that incur exclusively at the presence by ad-blockers and to look for potential leakages in information\footnote{e.g. \url{http://thehackernews.com/2016/11/web-of-trust-addon.html}}.
The domains which most frequently occur only in $\mathcal{E}$ are shown in Table~\ref{tab:exclusive}.
In many cases, we notice that $\mathcal{E}$ contains well-known ad-serving or visitor-tracking domains.
This clearly demonstrates that the tools are not perfect and are sometimes bypassed.

\begin{table}[]
\centering
\caption{Additional Domains loaded by Ad-blockers}
\label{tab:exclusive}
\begin{tabular}{c|c}
\hline
\textbf{Ad-blocker} & \textbf{Extra domains loaded}                                                            \\
\hline
AdBlock            & \texttt{mixpanel.com}, \texttt{stripe.com}                                                                                     \\
\hline
Adblock Plus       & {\tt quantserve.com},{\tt crwdcntrl.net}                                                                                   \\
\hline
Ghostery           & \begin{tabular}[c]{@{}l@{}}\texttt{selectmedia.asia}, \texttt{streamrail.com}\\ \texttt{adlooxtracking.com}\end{tabular} \\
\hline
Privacy Badger     & \texttt{cdnjs.com}  \\
\hline
uBlock             & Twitter widgets                                                                                                                   
\end{tabular}
\end{table}

For \texttt{AdBlock}, we notice that \texttt{mixpanel.com} appears in 5,125 $(17\%)$ of the webpages when using \texttt{AdBlock}, versus 170 $(<1\%)$ of original pages (vanilla mode). 
This is a Google Analytics-like website that tracks visits by users of \texttt{AdBlock}. 
Similarly, \texttt{stripe.com}, an online payments platform through which users can donate to \texttt{AdBlock} is added to 2,616 ($8\%$) of the webpages, compared to $40$ $(<0.1\%)$ with out ad-blocking. 

For {\tt Adblockplus}, we notice an increase for {\tt quantserve .com} (an audience-measurement website) from 14\% to 21\%
and for {\tt crwdcntrl.net} (another audience-measurement website) from 21\% to 29\%
For {\tt Ghostery}, we notice \texttt{selectmedia.asia} (a video-ad serving company) appears in 1,374 ($4.6\%$) webpages' requests, compared to 47 ($<0.1\%$) in the vanilla mode, \texttt{streamrail. com} (another video-ad serving company) 1,488 times ($5\%$) versus 63, and \texttt{adlooxtracking.com} (ad tracking company) 2,190 times ($7\%$) versus 240 ($<1\%$).
For {\tt PrivacyBadger}, \texttt{cdnjs.com} (a repository for open-source web-libraries) stands out with 2,130 webpages ($7\%$).
For {\tt uBlock} we did not witness any significant increase for suspicious domains; but we did notice an increase for Twitter widgets from 11\% to 20\% of the webpages.

Our preliminary findings indicate that though ad-blockers block external ads and third party trackers to varying degrees, they introduce various tracking services of their own. This is a novel finding and has potential impact on users privacy. We leave a detailed analysis on this for future work.




\spara{Conclusions}
This paper provides a first look at the performance and privacy aspects of popular ad-blocking tools. Based on our analysis, we conclude that (i) uBlock has the best performance, in terms of ad and third party tracker filtering, and least privacy tracking, (ii) The time to load pages is not necessarily faster when using adblockers, and (iii) this is partly the case due to additional trackers and libraries introduced by the adblocking tools.
As ad-blockers try to monetise their business by selling ads themselves, our datasets and findings could be used (i) by users, to get an idea on which adblocker to choose, from an increasingly competitive set of providers, and (ii) by researchers, as building blocks for further analysis on privacy and tracking behaviors by ad-blocking tools.

\spara{Acknowledgements.}
This work has been supported by the Academy of Finland project ``Nestor'' (286211) and the EC H2020 RIA project ``SoBigData'' (654024).

%% file: adblock.bbl

\begin{thebibliography}{00}


\ifx \showCODEN    \undefined \def \showCODEN     #1{\unskip}     \fi
\ifx \showDOI      \undefined \def \showDOI       #1{{\tt DOI:}\penalty0{#1}\ }
  \fi
\ifx \showISBNx    \undefined \def \showISBNx     #1{\unskip}     \fi
\ifx \showISBNxiii \undefined \def \showISBNxiii  #1{\unskip}     \fi
\ifx \showISSN     \undefined \def \showISSN      #1{\unskip}     \fi
\ifx \showLCCN     \undefined \def \showLCCN      #1{\unskip}     \fi
\ifx \shownote     \undefined \def \shownote      #1{#1}          \fi
\ifx \showarticletitle \undefined \def \showarticletitle #1{#1}   \fi
\ifx \showURL      \undefined \def \showURL       #1{#1}          \fi
\providecommand\bibfield[2]{#2}
\providecommand\bibinfo[2]{#2}
\providecommand\natexlab[1]{#1}
\providecommand\showeprint[2][]{arXiv:#2}

\bibitem[\protect\citeauthoryear{Barford, Canadi, Krushevskaja, Ma, and
  Muthukrishnan}{Barford et~al\mbox{.}}{2014}]%
        {barford2014adscape}
\bibfield{author}{\bibinfo{person}{Paul Barford}, \bibinfo{person}{Igor
  Canadi}, \bibinfo{person}{Darja Krushevskaja}, \bibinfo{person}{Qiang Ma},
  {and} \bibinfo{person}{S Muthukrishnan}.} \bibinfo{year}{2014}\natexlab{}.
\newblock \showarticletitle{Adscape: Harvesting and analyzing online display
  ads}. In \bibinfo{booktitle}{{\em WWW}}. \bibinfo{pages}{597--608}.
\newblock


\bibitem[\protect\citeauthoryear{Evans}{Evans}{2008}]%
        {evans2008economics}
\bibfield{author}{\bibinfo{person}{David~S Evans}.}
  \bibinfo{year}{2008}\natexlab{}.
\newblock \showarticletitle{The economics of the online advertising industry}.
\newblock \bibinfo{journal}{{\em Review of network economics\/}}
  \bibinfo{volume}{7}, \bibinfo{number}{3} (\bibinfo{year}{2008}).
\newblock


\bibitem[\protect\citeauthoryear{Falahrastegar, Haddadi, Uhlig, and
  Mortier}{Falahrastegar et~al\mbox{.}}{2014}]%
        {falahrastegar2014anatomy}
\bibfield{author}{\bibinfo{person}{Marjan Falahrastegar},
  \bibinfo{person}{Hamed Haddadi}, \bibinfo{person}{Steve Uhlig}, {and}
  \bibinfo{person}{Richard Mortier}.} \bibinfo{year}{2014}\natexlab{}.
\newblock \showarticletitle{Anatomy of the third-party web tracking ecosystem}.
\newblock \bibinfo{journal}{{\em arXiv preprint arXiv:1409.1066\/}}
  (\bibinfo{year}{2014}).
\newblock


\bibitem[\protect\citeauthoryear{Malloy, McNamara, Cahn, and Barford}{Malloy
  et~al\mbox{.}}{2016}]%
        {malloy2016ad}
\bibfield{author}{\bibinfo{person}{Matthew Malloy}, \bibinfo{person}{Mark
  McNamara}, \bibinfo{person}{Aaron Cahn}, {and} \bibinfo{person}{Paul
  Barford}.} \bibinfo{year}{2016}\natexlab{}.
\newblock \showarticletitle{Ad Blockers: Global Prevalence and Impact}. In
  \bibinfo{booktitle}{{\em Proceedings of the 2016 ACM on Internet Measurement
  Conference}}. \bibinfo{pages}{119--125}.
\newblock


\bibitem[\protect\citeauthoryear{Nithyanand et~al\mbox{.}}{Nithyanand
  et~al\mbox{.}}{2016}]%
        {nithyanand2016ad}
\bibfield{author}{\bibinfo{person}{Rishab Nithyanand} {and}
  \bibinfo{person}{others}.} \bibinfo{year}{2016}\natexlab{}.
\newblock \showarticletitle{Ad-Blocking and Counter Blocking: A Slice of the
  Arms Race}.
\newblock \bibinfo{journal}{{\em arXiv:1605.05077\/}} (\bibinfo{year}{2016}).
\newblock


\bibitem[\protect\citeauthoryear{PageFair}{PageFair}{2015}]%
        {pagefair2015report}
\bibfield{author}{\bibinfo{person}{PageFair}.} \bibinfo{year}{2015}\natexlab{}.
\newblock \showarticletitle{The 2015 Ad Blocking Report.}
\newblock \bibinfo{journal}{{\em
  https://blog.pagefair.com/2015/ad-blocking-report/\/}}
  (\bibinfo{year}{2015}).
\newblock


\bibitem[\protect\citeauthoryear{PageFair}{PageFair}{2016}]%
        {pagefair2016report}
\bibfield{author}{\bibinfo{person}{PageFair}.} \bibinfo{year}{2016}\natexlab{}.
\newblock \showarticletitle{The 2016 Mobile Ad Blocking Report.}
\newblock \bibinfo{journal}{{\em
  https://pagefair.com/blog/2016/mobile-adblocking-report/\/}}
  (\bibinfo{year}{2016}).
\newblock


\bibitem[\protect\citeauthoryear{Pujol, Hohlfeld, and Feldmann}{Pujol
  et~al\mbox{.}}{2015}]%
        {pujol2015annoyed}
\bibfield{author}{\bibinfo{person}{Enric Pujol}, \bibinfo{person}{Oliver
  Hohlfeld}, {and} \bibinfo{person}{Anja Feldmann}.}
  \bibinfo{year}{2015}\natexlab{}.
\newblock \showarticletitle{Annoyed Users: Ads and Ad-Block Usage in the Wild}.
  In \bibinfo{booktitle}{{\em IMC}}. \bibinfo{pages}{93--106}.
\newblock


\bibitem[\protect\citeauthoryear{Schelter and Kunegis}{Schelter and
  Kunegis}{2016}]%
        {schelter2016ubiquity}
\bibfield{author}{\bibinfo{person}{Sebastian Schelter} {and}
  \bibinfo{person}{J{\'e}r{\^o}me Kunegis}.} \bibinfo{year}{2016}\natexlab{}.
\newblock \showarticletitle{On the Ubiquity of Web Tracking: Insights from a
  Billion-Page Web Crawl}.
\newblock \bibinfo{journal}{{\em arXiv preprint arXiv:1607.07403\/}}
  (\bibinfo{year}{2016}).
\newblock


\bibitem[\protect\citeauthoryear{Walls, Kilmer, Lageman, and McDaniel}{Walls
  et~al\mbox{.}}{2015}]%
        {walls2015measuring}
\bibfield{author}{\bibinfo{person}{Robert~J Walls}, \bibinfo{person}{Eric~D
  Kilmer}, \bibinfo{person}{Nathaniel Lageman}, {and}
  \bibinfo{person}{Patrick~D McDaniel}.} \bibinfo{year}{2015}\natexlab{}.
\newblock \showarticletitle{Measuring the Impact and Perception of Acceptable
  Advertisements}. In \bibinfo{booktitle}{{\em IMC}}.
  \bibinfo{pages}{107--120}.
\newblock


\bibitem[\protect\citeauthoryear{Wills and Uzunoglu}{Wills and
  Uzunoglu}{2016}]%
        {wills2016ad}
\bibfield{author}{\bibinfo{person}{Craig~E Wills} {and}
  \bibinfo{person}{Doruk~C Uzunoglu}.} \bibinfo{year}{2016}\natexlab{}.
\newblock \showarticletitle{What Ad Blockers Are (and Are Not) Doing}. In
  \bibinfo{booktitle}{{\em HotWeb}}. \bibinfo{pages}{72--77}.
\newblock


\bibitem[\protect\citeauthoryear{Yu, Macbeth, Modi, and Pujol}{Yu
  et~al\mbox{.}}{2016}]%
        {yu2016tracking}
\bibfield{author}{\bibinfo{person}{Zhonghao Yu}, \bibinfo{person}{Sam Macbeth},
  \bibinfo{person}{Konark Modi}, {and} \bibinfo{person}{Josep~M Pujol}.}
  \bibinfo{year}{2016}\natexlab{}.
\newblock \showarticletitle{Tracking the Trackers}. In \bibinfo{booktitle}{{\em
  WWW}}. \bibinfo{pages}{121--132}.
\newblock


\end{thebibliography}
